\documentclass[10pt,conference]{IEEEtran}
\IEEEoverridecommandlockouts
\usepackage{cite}
\usepackage{amsmath,amssymb,amsfonts}
\usepackage{algorithmic}
\usepackage{graphicx}
\usepackage{textcomp}
\usepackage{xcolor}
\usepackage[T1]{fontenc}
\def\BibTeX{{\rm B\kern-.05em{\sc i\kern-.025em b}\kern-.08em
    T\kern-.1667em\lower.7ex\hbox{E}\kern-.125emX}}
\begin{document}

\title{From Ambiguous Tickets to Ranked Hypotheses: Multi-Agent Root-Cause Analysis for Autonomous Heavy-Vehicle Software
}

\author{
\IEEEauthorblockN{
Marten Seegers\IEEEauthorrefmark{1},
Maximilian Kneissl\IEEEauthorrefmark{2},
Enrique Aguayo-Lara\IEEEauthorrefmark{2},
Chih-Hong Cheng\IEEEauthorrefmark{1}\IEEEauthorrefmark{3}
}
\IEEEauthorblockA{
\IEEEauthorrefmark{1}
\textit{Carl von Ossietzky University of Oldenburg}, Oldenburg, Germany
}
\IEEEauthorblockA{
\IEEEauthorrefmark{2}
\textit{Volvo Autonomous Solutions}, Gothenburg, Sweden
}
\IEEEauthorblockA{
\IEEEauthorrefmark{3}
\textit{Chalmers University of Technology}, Gothenburg, Sweden
}
\IEEEauthorblockA{
\{marten.seegers, chih-hong.cheng\}@uni-oldenburg.de, \{maximilian.kneissl, enrique.aguayo\}@volvo.com
}
}

\maketitle

\begin{abstract}
Autonomous haulage systems run safety-critical software on driverless mining trucks, with no onboard driver to detect or mitigate abnormal behavior. Every failure in testing or operation must therefore be traced to its root cause to keep the safety case valid and limit downtime. An investigation often starts from a sparse ticket containing an approximate timestamp, sometimes a suspected cause, and a multimodal vehicle recording of several gigabytes. We present a multi-agent root-cause analysis (RCA) workflow that turns this input into a telemetry-grounded event chain, ranked causal hypotheses, and candidate code locations. A suspected cause in the ticket, which could anchor the analysis on a false theory, goes to a single LLM analyst in a separate agent session. Non-neural tools select the relevant telemetry once, before analysis, and an LLM condenses it into shared evidence artifacts; this more than halved the cost per incident compared with per-analyst querying. Specialist analyst agents then apply established RCA techniques in parallel, including fault tree analysis, Kepner-Tregoe analysis, and call-chain tracing, and hypotheses that several LLM analysts reach independently rank higher. We evaluated the workflow on 12 held-out industrial incidents against a single agent with the same model and tools. It located the faulty file more often (10 vs. 8 cases) at a higher cost. Two human domain experts accepted a top-three hypothesis in 4 and 9 cases, respectively. The gap, which two LLM judges did not resolve, shows that rating causal explanations is itself a hard evaluation problem.
\end{abstract}


\section{Introduction}
Autonomous heavy vehicles are deployed in mining and construction, where Autonomous Haulage Systems (AHS) promise greater safety and operational efficiency by removing personnel from hazardous environments and reducing dependence on human shift constraints~\cite{du2025industrial}. Volvo Autonomous Solutions (V.A.S.) operates such a system commercially at European customer sites~\cite{vas2026threemt}. Software failures in these safety-critical cyber-physical systems can interrupt production and threaten human safety and property. Therefore, diagnosis requires reasoning across perception, localization, planning, control, state management, and the physical environment, using bug tickets, telemetry, source code, and configuration artifacts. This root-cause analysis (RCA) is cognitively demanding and time-consuming~\cite{chen2025empirical}.

In this paper, we investigate incidents recorded in a sparse, possibly inaccurate ticket and a single multimodal vehicle recording. Recent work applies large language models (LLMs) and agents to diagnose cloud outages~\cite{ahmed2023recommending,chen2024automatic,roy2024exploring,li2025coca}, while autonomous-driving studies rely on re-running accidents or injecting faults in simulation~\cite{feng2024rocas,lu2024diavio}. We instead want to design an agent system that performs analysis on the static evidence available after an incident occurred. In our early experiments, a single agent struggled when ticket timestamps were off, latched onto causes suggested by the person who filed the ticket, and settled too quickly on the first plausible explanation. In controlled evaluations of LLM based cloud RCA, Riddell et al.~\cite{riddell2026stalled} found that anchoring on a single hypothesis without exploring alternatives and failing to revise a claim when new evidence contradicts it are among the strongest predictors of an incorrect diagnosis. Our specific question is how to organize LLM based analysis to localize an incident in gigabytes of telemetry and identify plausible defects without letting the human reporter's theory or an early hypothesis dominate.

For this purpose, we develop an evidence-oriented multi-agent RCA workflow for the AHS, which produces a telemetry-grounded event chain, ranked causal hypotheses, candidate affected code locations and next investigation steps.
The workflow rests on four design choices. First, the human reporter's causal theory is removed from the shared symptom description and given to one LLM analyst. Because every LLM analyst runs in its own agent session, the others never see it. Second, the expensive work on the recording happens once, before the LLM analysts start. This is mainly done with non-neural tools, which more than halved the median cost per incident. Third, different LLM analysts work in parallel, each using established but different RCA techniques, among them fault tree analysis~\cite{vesely1981fault} and Kepner-Tregoe problem analysis~\cite{kepner1965rational}, and a hypothesis that several analysts reach independently ranks higher. Fourth, shared domain context files and explicit checks of physical and temporal consistency help LLM analysts interpret evidence within vehicle and environmental constraints.

The workflow is evaluated on 12 held-out historical incidents, each isolated at its pre-fix code state. We compare code-location accuracy with a single agent using the same model, tools, and repository access. Two human domain experts and two LLM judges rate the workflow's three top-ranked hypotheses and its event chain against the known fix, and we trace accepted hypotheses to their analysts. The workflow found the faulty file among its three highest-ranked hypotheses in 10 of 12 cases, compared with 8 for the single agent, at a median cost of US\$8.69 versus US\$0.94 per incident. In eight of the nine cases with an accepted hypothesis, the highest-ranked accepted hypothesis had contributions from multiple LLM analysts. Two cases illustrate the tradeoff in handling human reporter theories: the baseline followed a misleading theory in one, while only the dedicated LLM analyst produced a fully correct hypothesis in another. Explaining the failure remained harder to assess than locating the code. The two experts accepted a top-three hypothesis in 9 and 4 cases, respectively, and three cases with matching code locations had no accepted explanation. Event chains still helped, with both experts rating them as narrowing the search to the right area or better in 8 cases, including two where neither accepted a hypothesis. These observations are limited to 12 incidents from one industrial codebase.

We summarize our contribution as follows:
\begin{itemize}
    \item A cost-effective industrial multi-agent RCA workflow that turns ambiguous tickets and vehicle telemetry into auditable event chains, ranked causal hypotheses, and candidate code locations.
    \item Four design choices for LLM-based RCA: (i)~bias-isolated intake via separate agent sessions, (ii)~evidence construction and control by non-neural tools, (iii)~corroborating specialists using established RCA techniques, and (iv)~shared domain knowledge with physical and temporal consistency checks.
    \item A leakage-controlled evaluation on 12 held-out industrial incidents against a single-agent baseline, rated by two human experts and two LLM judges, confirming the effectiveness of our proposed approach.
\end{itemize}

\section{Related Work} 

\subsection{LLM-Based Root-Cause Analysis}
Ahmed et al.~\cite{ahmed2023recommending} were among the first to evaluate language models on large-scale cloud incidents. They found that automatic text-similarity metrics are useful only for comparing models against one another, while judging actual diagnostic correctness still requires human experts. Subsequent work gave models more structure and tool access. RCACopilot~\cite{chen2024automatic} runs alert-specific handlers to gather diagnostic data and predicts a root-cause category by retrieving similar past incidents. Roy et al.~\cite{roy2024exploring} showed that a ReAct agent equipped with retrieval and diagnostic tools matches retrieval-based methods while hallucinating less, and COCA~\cite{li2025coca} traces execution paths starting from log statements mentioned in bug reports. Multi-agent systems such as RCAgent~\cite{wang2024rcagent} and Flow-of-Action~\cite{pei2025flow} split the investigation across several tool-using agents, with Flow-of-Action guiding agents through encoded standard operating procedures.

Diagnosing failures in our autonomous haulage trucks differs from cloud RCA in the evidence available. Cloud systems can be queried live and draw on thousands of similar past outages. A vehicle incident however leaves behind a single recording of a physical run, and failures are too varied and rare to solve by retrieving past tickets.

\subsection{Failures of Agentic Reasoning}
Riddell et al.~\cite{riddell2026stalled} analyzed 48{,}000 simulated diagnostic runs and grouped agent reasoning mistakes into 16 failure modes. They found that once an agent forms an early theory or adopts a hint from the prompt it rarely changes course even when later tool outputs contradict it. This finding directly motivated our decision to hide the human reporter's suspected cause from all but one analyst. Cemri et al.~\cite{cemri2026multi} analyzed 150 multi-agent execution traces by hand to define 14 failure modes, showing that poorly coordinated agents often repeat work, drift off task, or add little over a single agent. To avoid some of these failure modes, our workflow gives each analyst a different fixed RCA method, a shared set of extracted log evidence, and a strict output schema checked by deterministic code.

\subsection{Log Analysis and Evidence Construction}
Classical log-anomaly detectors such as DeepLog~\cite{du2017deeplog} learn what normal log sequences look like from training data, so their accuracy drops when software updates introduce unfamiliar log messages~\cite{liu2024interpretable}. Newer methods use LLMs to parse log templates more efficiently, as in LILAC's adaptive parsing cache~\cite{jiang2024lilac}, or add explanations to anomaly detection, as in LogGPT~\cite{qi2023loggpt}. However, these tools are designed for uniform text logs and output either parsed templates or anomaly flags. Vehicle recordings combine text logs with continuous sensor signals, state transitions, and 3D vehicle poses. Our preprocessing stage pulls these different streams together into a single time-aligned event chain that clearly distinguishes raw measurements from agent interpretations.

\subsection{Failure Analysis for Autonomous Driving}
Many existing diagnostic tools for autonomous driving rely on simulation. ROCAS~\cite{feng2024rocas} narrows down root causes by repeatedly re-running a scenario while mutating software and physical parameters, and DiaVio~\cite{lu2024diavio} diagnoses safety violations produced by scenario-based simulation testing. Both require a simulated, replayable scenario, whereas our goal is a workflow that needs no scenario reconstruction or re-execution and works from the incident recording alone. Empirical bug studies in autonomous driving and robotics inform our design and evaluation instead. Garcia et al.~\cite{garcia2020comprehensive} classify 499 bugs in Apollo and Autoware into 13 root causes across components from perception to control, and Timperley et al.~\cite{timperley2024robust} study 221 bugs across seven ROS systems in the ROBUST dataset, finding that robotics failures often span multiple communicating components even when the fix is localized to a single file. Both studies motivate cross-component reasoning, and we follow ROBUST's benchmark practice of isolating each historical bug at the parent commit of its fix.

\section{Operational Setting and Requirements}\label{sec:setting}
We target RCA for incidents in the onboard autonomous driving software of the V.A.S. AHS. The AHS is a distributed cyber-physical system comprising onboard autonomy, vehicle platform control, perception and localization, digital site maps, site infrastructure, and cloud-based fleet management.
The onboard driving software sits at the center of the system. It receives missions from fleet management and combines them with localization, perception, and vehicle-state data to plan trajectories, control the truck, monitor system health, and switch driving states. Because these modules depend on one another, a problem noticed in the driving software may actually be caused by upstream sensors, timing delays between modules, bad configuration values or external services. The mining environment adds complexity as well: road conditions, weather, visibility, vehicle load, and repeated use of haul routes can affect behavior and complicate symptom interpretation. 

Incidents assigned to the driving software team come from simulation, test tracks, or commercial customer sites, and the initial bug tickets are often brief.  A typical ticket reports a visible symptom — for example, that the truck stopped unexpectedly or refused to change state — without describing what happened beforehand. Sometimes the person reporting the incident also includes a theory about what the cause might be. Experienced operators sometimes spot operational issues that are not obvious in the logs, but their guesses can also be wrong and might lead the investigation the wrong way. The reported timestamp can be unreliable as well. Because many tickets are written in the field after the truck has already stopped, the time noted in the ticket can be several minutes after the fault actually started, whereas automated log analysis needs the fault window narrowed down to within a few seconds.

The main source of evidence for locating the failure is the recording from the vehicle run, which bundles sensor signals, state transitions, and text logs from across the AHS into a single container. Even a few minutes of driving can produce gigabytes of data, which is far too large and costly to feed directly into an LLM, especially since language models struggle to find key facts buried in very long inputs~\cite{liu2024lost}. At the same time, looking only at the driving software's text log can miss faults triggered by other subsystems. The workflow therefore needs a preprocessing step that locates the relevant time window, captures the state of all subsystems right before and during the failure, and condenses the recording into a short timeline that both agents and engineers can read.

These practical constraints lead to five requirements for our workflow:
\begin{enumerate}
    \item \emph{Cross-subsystem analysis:} It must trace how faults develop across neighboring subsystems, timing boundaries, and physical site conditions.
    \item \emph{Robust ticket intake:} It must verify the reported timestamp against the logs and isolate the reporter's causal guesses from observed symptoms.
    \item \emph{Compact, traceable evidence:} It must condense gigabytes of raw telemetry into short summaries while preserving evidence sources.
    \item \emph{Multiple ranked hypotheses:} Because several explanations often fit the initial symptoms, the workflow should explore alternatives in parallel and return a short ranked list of hypotheses. Engineers (or automated tests) can then check the top candidates in order.
    \item \emph{Standalone intermediate summaries:} It must save and visualize the extracted timeline and evidence so that, even when all hypotheses are wrong, engineers still get a clear overview of the incident to start their own debugging.
\end{enumerate}
\begin{figure*}
    \centering
    \includegraphics[width=0.9\linewidth]{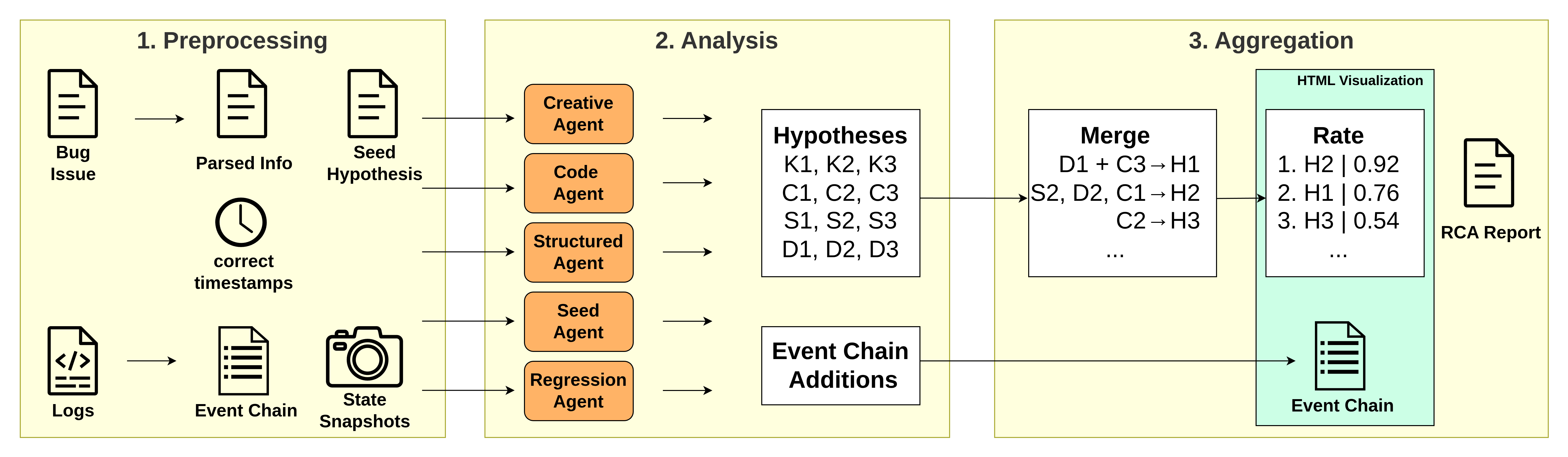}
    \caption{The three phases of the RCA workflow. Hypothesis IDs name the producing LLM analyst (S structured, K creative, C code, D seed evaluator).}
    \label{fig:architecure_workflow}
\end{figure*}
\section{System Overview}
Figure~\ref{fig:architecure_workflow} shows the three phases of the workflow: preprocessing, parallel analysis, and aggregation. A main orchestrator agent interacts with the user, launches subagents defined by prompt files, and collects their outputs, while code-based tools handle log queries, code searches, and schema checks.

\subsection{Preprocessing}
First, the intake step reads the bug ticket and creates a structured incident record with the incident metadata and a plain description of the symptoms. If the ticket contains a specific guess about what caused the failure, that guess is pulled out and saved separately as a \emph{seed hypothesis}. The description of the failure is rewritten in neutral terms that contain only the observed symptoms.

Next a timestamp-validation step locates the failure in the recording. Starting from the timestamp given in the ticket, the logs are searched for two anchors: the last moment the system was operating normally and the moment the first error appeared. These anchors are normalized to UTC and define the evidence window for later stages.

Once the incident time is identified, an extraction step builds two shared evidence files from the recording. The first is a \emph{state snapshot report}, which lists the status of all subsystems at the two time anchors. The second is an \emph{event chain}, a chronological timeline of key log messages, state transitions, signal anomalies, and vehicle context across the incident window.

\subsection{Parallel analysis}

The orchestrator then launches the specialized LLM analysts of Table~\ref{tab:analysts} in parallel, each in its own agent session. Every analyst receives the neutral incident record, the state snapshot, and the event chain, and can query the code and logs for further details. An analyst that notices a relationship the extraction missed can propose an addition to the event chain, tagged as inferred. Four analysts run on every incident. The fifth, the seed evaluator, runs only when a suspected cause was identified in the ticket.
\begin{table}[t]
\centering
\caption{Specialized LLM analysts of the parallel-analysis phase.}
\label{tab:analysts}
\footnotesize
\renewcommand{\arraystretch}{1.2}
\begin{tabular}{@{}p{0.24\linewidth} p{0.7\linewidth}@{}}
\hline
Analyst & Role \\
\hline
Structured & Reconstructs the timeline and the causal dependencies with causal graphs and fault tree analysis~\cite{vesely1981fault} \\
Creative & Probes alternative explanations using 5 Whys~\cite{ohno1988toyota}, Kepner-Tregoe~\cite{kepner1965rational}, and Ishikawa diagrams~\cite{ishikawa1976guide} \\
Code & Maps observed behavior to implementation paths, affected modules, and timing or state-handling issues by tracing message flows and call chains \\
Regression & Inspects recent code history for plausible introducing changes; abstains unless it finds evidence of a regression \\
Seed evaluator & Evaluates the human reporter's explanation in isolation, searching for supporting and contradicting evidence \\
\hline
\end{tabular}
\end{table}

\subsection{Aggregation and reporting}
In the final phase, the orchestrator combines overlapping hypotheses from different analysts into a single pool. A separate LLM assessor then rates each merged hypothesis on five quality criteria, and a fixed scoring formula combines those ratings with the number of analysts that found each hypothesis to produce the final ranking. The workflow outputs a readable RCA report along with files for all intermediate steps (the incident record, state snapshot, event chain, individual analyst outputs, and ranked hypothesis pool) as well as an interactive HTML visualization.

\section{Key Design Choices}\label{sec:design}

\subsection{Bias-Isolated Intake}\label{sec:intake}
A ticket's causal theory may carry domain knowledge, but it can also anchor every LLM analyst on one narrative~\cite{riddell2026stalled}. Therefore we do not discard the theory, but restrict who sees it. While all LLM analysts receive the neutral symptom description produced at intake, only the seed evaluator receives the seed hypothesis, and it must search for both supporting and contradicting evidence. Its result then competes with the other hypotheses without special status. The separation works only because each analyst runs in its own session with its own context window, so information can reach one analyst without reaching the others.

\subsection{Non-Neural Tools for Cost and Control}\label{sec:tools}
In an earlier prototype of our system, each LLM analyst queried the recording independently through telemetry tools, repeating much of the expensive traversal. That version had a median cost of 2{,}109 credits\footnote{GitHub Copilot AI credits, computed from the input, cached, and output tokens of a run at fixed per-model prices. One credit currently converts to US\$0.01~\cite{github2026copilot}.} per incident. The current workflow instead centralizes this work before analysis, using mostly non-neural tools. For timestamp validation, a keyword search collects candidate error events from the logs and presents them to the validator agent. For extraction, statistical and rule-based anomaly detection select the relevant signals and events, substantially shrinking the model input. A separate agent session then condenses this raw evidence into readable entries and puts it in context, producing the state snapshot report and the event chain. During enrichment it can still perform additional queries of the logs; due to the scripted initial extraction however, the number of queries needed is significantly smaller. On our benchmark, the workflow's median cost is 869 credits per incident (Section~\ref{sec:results}), which is less than half that of the earlier version. Shared evidence gives LLM analysts the same starting point and ties hypotheses to inspectable artifacts.

Non-neural tools also keep the agents in check. Any operation that can be specified unambiguously runs as deterministic code instead of being run through a model. Agents create and edit artifacts through a deterministic artifact store. The store validates each hypothesis against the schema of Section~\ref{sec:analysts} and records which LLM analysts contributed to it. During synthesis, the store compares all submitted hypotheses with simple text-similarity measures and suggests candidate merges. The orchestrator merges candidates describing the same mechanism, combining evidence, predictions, and code locations, while distinct mechanisms remain separate. During ranking LLM analysts do not score their own hypotheses, because confidence values obtained from language models are poorly calibrated~\cite{xiong2024can}. A separate LLM assessor instead rates the merged hypotheses comparatively on named ordinal dimensions, namely explanatory power, evidence balance, mechanism specificity, discrimination from alternatives, and spatial-temporal consistency. The store scores and ranks the pool using these ratings and the number of contributing analysts. Restricting the assessor to ordinal ratings limits the influence of known LLM judge biases such as position preference and self-preference~\cite{zheng2023judging, panickssery2024llm}.

\subsection{Specialized LLM analysts with Established RCA Techniques}\label{sec:analysts}
For a failure incident in our setting several explanations can seem plausible at first sight, and agentic reasoning has a tendency to stop at the first that fits~\cite{riddell2026stalled}. For that reason we split the analysis among specialists that work in separate agent sessions and approach the incident with different established RCA techniques (Table~\ref{tab:analysts}). This widens the search and also supplies a ranking signal unavailable to a single agent. When LLM analysts that worked differently arrive at the same mechanism, the merged hypothesis records their agreement, and the ranking counts it as corroboration (Section~\ref{sec:tools}).

The structured analyst decomposes the incident into a detailed timeline and causal graph, checking each link in the code by tracing a logged topic back to the module that publishes it. It then builds a fault tree~\cite{vesely1981fault} with the reported symptom as the top event and follows AND and OR gates down to basic events like lowest-level code conditions or signal values. Each path from the top event to a basic event becomes a candidate hypothesis. The creative analyst on the other hand follows a more liberal approach to find causes this decomposition can miss. It applies 5 Whys~\cite{ohno1988toyota} to follow one effect back through repeated why-questions, Kepner-Tregoe problem analysis~\cite{kepner1965rational} to contrast where and when the problem occurred with where and when it could have but did not, and an Ishikawa diagram~\cite{ishikawa1976guide} to sort candidate causes into categories so that none is overlooked. The code analyst instead focuses on the implementation. It follows message and event flows from logged topics to the modules that produce them, across module boundaries where needed, and traces the call chains of the functions that decide the observed behavior.

For comparability, LLM analysts use a fixed machine-readable schema with fields for a single falsifiable causal statement, a mechanism category, supporting evidence and counter-evidence with source, predictions, causal links between event chain entries, candidate code locations, and open questions. At least one counter-evidence entry is mandatory, so analysts must check conditions that weaken their explanation or explicitly state none were found. Shared fields allow findings to be merged field by field rather than combined as free text.

\subsection{Domain Context and Physical Reasoning}
Evaluating whether a fault mechanism is plausible requires checking both the physical scene and the timing of asynchronous updates. During preprocessing, the telemetry server extracts driving context and subsystem snapshots for localization, perception, control, health, and mission state, placing the truck's pose, heading, speed, planned trajectory, obstacle regions, and message timestamps into the shared state snapshot and event chain. Alongside these incident-specific files, every agent session automatically loads files that document the stack's module interfaces, expected event meanings, known benign operational warnings, and diagnostic rules of thumb. Known benign warnings are tagged in the event chain during extraction so that analysts do not mistake routine operational noise for a symptom.

Physical and temporal checks are then enforced in two places. First, analysts must verify that their proposed cause fits the recorded geometry and update order, such as checking whether an obstacle lay within sensor coverage from the truck's pose or whether a signal was published before a consumer module acted on it, and record the outcome in the hypothesis schema's supporting or mandatory counter-evidence fields. Second, during synthesis, the LLM assessor scores each merged hypothesis on spatial-temporal consistency, lowering the rank of explanations that conflict with the recorded physical state or timeline.
\section{Evaluation}

Two difficulties shape our evaluation on historical incidents. First, the output of the workflow is free text, and the correctness of a root cause cannot be confirmed by measuring semantic similarity to a ground truth reference text~\cite{ahmed2023recommending}. Second, defining the ground truth requires manual analysis of solved incidents' fixes, which limits the size of the benchmark. We therefore use 12 incidents chosen for diversity (Section~\ref{sec:benchmark}) and two kinds of measures, code-location accuracy as a proxy for correctness and ratings by human domain experts who are familiar with the fixes. The experts judge whether each hypothesis identifies the defect. They separately rate event chain usefulness and clutter, as this analysis artifact should guide engineers without burying relevant events, even when no hypothesis is correct (Section~\ref{sec:setting}). Given that the time of experts is limited, two LLM judges rate the same 60 items and we report their agreement with the experts.

To evaluate our system, we ask three research questions:

\textbf{RQ1:} Does multi-agent decomposition improve code-location accuracy over a single agent with the same model, tools, and repository access?

\textbf{RQ2:} How often does one of the workflow's three top-ranked
hypotheses identify the defect, and how useful are its event chains, when rated against the known fix by domain experts and LLM judges?

\textbf{RQ3:} How do the specialist analysts and the bias-isolated intake contribute to the workflow's correct hypotheses?

\subsection{Benchmark Set Diversity}\label{sec:benchmark}

\begin{figure}
    \centering
    \includegraphics[width=1\linewidth]{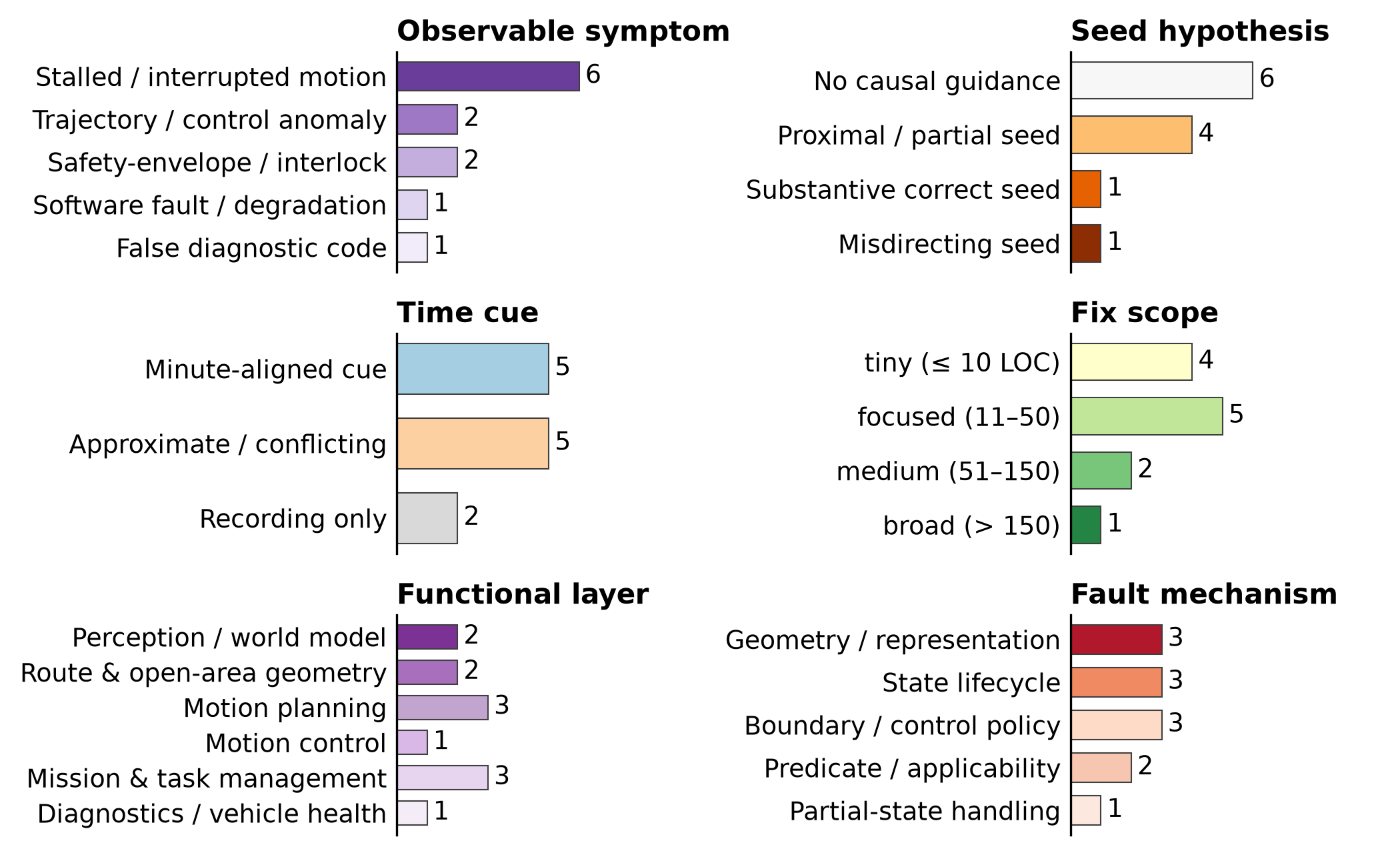}
    \caption{Coverage of the 12 benchmark cases along six dimensions: observable symptom, functional layer, fix scope, seed hypothesis, time cue, and fault mechanism.}
    \label{fig:benchmark_diversity}
\end{figure}

We evaluate on 12 solved historical incidents for which the relevant vehicle recording was available. While it cannot represent the incident population this purposive maximum-variation sample still tests heterogeneous conditions. Fig.~\ref{fig:benchmark_diversity} shows its coverage along six dimensions. \emph{Observable symptom}, \emph{functional layer}, and \emph{fix scope} describe the observable outcome, the faulty part of the autonomy stack, and the size of the gold patch. \emph{Seed hypothesis} is the causal guidance the ticket exposes, from none to an essentially correct hint, and \emph{time cue} is the temporal guidance quality, from minute-level alignment to conflicting timestamps or recording only. Both stress explicit workflow steps. \emph{Fault mechanism} is the class of defect, such as a geometry error, a state-lifecycle defect, a predicate error, or a boundary or control-policy issue. It varies independently of the observable symptom. Two state-lifecycle defects in mission management stall the vehicle, whereas a third bypasses a rollover interlock.

\subsection{Leakage-Controlled Evaluation Environment}\label{sec:leakage}
A benchmark of historical incidents risks leakage through post-incident artifacts like fixes, later history, developer discussions, or ground-truth annotations. We therefore isolate each case in a repository snapshot immediately before its fix. The workflow can inspect the code and a bounded portion of earlier development history, reflecting information available to engineers investigating a regression. All later history, including the fix and diagnosis, is unavailable. Prompts, domain context files, and heuristics were designed and tuned on a smaller separate set of incidents, and no benchmark case was used during tuning.

\subsection{Single-Agent Baseline}
The baseline is a single agent with one prompt and the same telemetry
tools, domain context files, and repository snapshot as the workflow. These tools include event chain and state snapshot extraction, so the baseline can construct equivalent evidence artifacts. Both configurations use Claude Sonnet 5 with high thinking effort and a 1M-token context window. The baseline has no orchestrator, specialized LLM analysts, seed evaluation, or synthesis stage, and it cannot dispatch subagents. It outputs an event chain and exactly three self-ranked hypotheses, each with a statement and affected code locations. We do not require the evidence and provenance fields of the workflow's hypothesis schema, since they would reintroduce part of the architecture under test.

\subsection{Code-Location Accuracy}
Code-location accuracy is an RCA quality proxy that can be scored against the gold patch without rater judgment. Although it does not establish mechanism correctness, naming the correct module, file, or function narrows the search to the faulty region. Ground truth combines the non-test files changed by the gold patch with a list of expected locations. We score at module, file, and function level. Function names come from the hunk headers of the patch. Predictions are the affected code locations of the top-1 and top-3 hypotheses, ordered by the workflow's ranking. Paths are normalized before matching, and a predicted directory counts only at module level.

\subsection{Expert Ratings}
Two engineers from the onboard driving software team rated the workflow output for all 12 cases. They applied a fixed codebook against a known fix. For each case, raters read the ticket and gold patch, rated event chain usefulness and clutter, then rated each top-three hypothesis for correctness (Table~\ref{tab:rubric}), which yields 60 items per rater. Case order was randomized per rater, and hypothesis order was shuffled within each case. Raters saw only the statement and affected code of each hypothesis, while its score, rank, and provenance remained hidden, so that the workflow's ranking can be tested against the expert ratings. Raters worked asynchronously and independently.

All anchors describe observable artifact properties. A single correctness scale rates the defect and the mechanism together. Usefulness reflects the next step prompted by the event chain, from going straight to the faulty component (5) to being sent in the wrong direction (1), and clutter ranges from almost no irrelevant events (5) to almost only irrelevant events (1).

\begin{table}[t]
\centering
\caption{Correctness scale for hypotheses. A not-applicable response is also offered.}
\label{tab:rubric}
\footnotesize
\renewcommand{\arraystretch}{1.3}
\begin{tabular}{@{}c p{0.88\linewidth}@{}}
\hline
\multicolumn{2}{@{}p{\linewidth}@{}}{\textbf{Hypothesis correctness.}
\emph{Compared with the ground truth, does this hypothesis identify the
defect and explain how it produced the reported symptom?}} \\
\hline
5 & Same defect and same mechanism; an engineer following this would
write the gold fix \\
4 & Same defect; mechanism incomplete or partly wrong; the fix would
still land in the right place \\
3 & Names a real contributing factor or the correct faulting subsystem,
but not the defect itself \\
2 & Names a plausible but wrong defect, or restates the symptom as its own
cause \\
1 & Unrelated to the observed failure \\
\hline
\end{tabular}
\end{table}
\subsection{LLM judges}
Two LLM judges from different model families, GPT-5.6 Terra and Claude
Sonnet 5, rated the same 60 items using the experts' rubric, blinded
content, and presentation order. Each judge rated each
item once. Claude Sonnet 5 also generated
the hypotheses under evaluation, and language models can
favor their own outputs~\cite{panickssery2024llm}. We report the LLM judges
separately. 

\subsection{Analysis}
The analysis plan was fixed before rating began. With 12 cases, all results are descriptive. For RQ1, we report top-1 and top-3 accuracy per granularity and show the per-case outcomes of workflow and baseline side by side.

For RQ2, a hypothesis is correct if rated 4 or higher as this means the fix would land in the right place (Table~\ref{tab:rubric}). A case is solved for a rater if any top-three hypothesis is correct. This mirrors the sequential verification described in Section~\ref{sec:setting}. A correct hypothesis at rank 2 or 3 still solves the case but requires more verification, so we report solved cases and the first accepted rank per rater and judge. We also report the distributions of event chain usefulness and clutter. For the expert pair and each expert--judge pair, we report ordinal Krippendorff's $\alpha$, exact agreement, and the number of cases with matching solved status. Exact agreement complements $\alpha$, which can be low when ratings cluster at one end of the scale even if raw agreement is high.

For RQ3, we look at which specialist analysts produced the highest-ranked accepted hypothesis in each solved case—the hypothesis where an engineer checking the list in order would stop (if the two experts accepted different ranks, we take the higher-ranked one). For each specialist role, we count how many of these accepted hypotheses it contributed to and how often it was the only analyst to find that hypothesis. Because code-location accuracy is available for all 12 cases, we also run a leave-one-out check on code localization: for each analyst role, we drop every top-three hypothesis that analyst helped produce and check whether the best remaining code match in each case drops to a coarser level (for example, from function level down to file level). This assumes the worst case, in which removing one analyst contributor leads to the hypothesis being incomplete or ranked significantly lower.

\section{Results}\label{sec:results}

\subsection{RQ1: Code-Location Accuracy}
Table~\ref{tab:per_case} shows the per-case outcomes. The workflow matched the gold patch more often than the baseline at every granularity and both cut-offs. At top-3 it located the faulty module in 11 of 12 cases, the file in 10, and the function in 6, against 9, 8, and 4 for the baseline. At top-1, counts were 9, 8, and 5 against 6, 6, and 2.

The top-3 outcomes were better than the baseline in four cases (C01, C02, C09, C10), worse in two (C05, C08), and equal in six. The baseline found no correct location in C01, C02, and C10, where the workflow reached the correct file at ranks 2 and 3 (C01, C02) and the correct function at rank 1 (C10). Every case missed by the workflow's first hypothesis was also missed by the baseline's, while the workflow more often ranked its best match first. Its first hypothesis reached its best-of-3 granularity in 8 of 11 cases with any match, the baseline's in 5 of 9, and the baseline's second-ranked hypotheses matched in more cases than its first (8 against 6).

The workflow consumed a median of 868.8 credits per case (range 614.0--976.9), or US\$8.69, and the baseline 94.2 credits (69.5--123.5), or US\$0.94. The per-case ratio had median 8.5, ranging from 5.7 (C12) to 13.4 (C09).

\begin{table}[t]
\centering
\caption{Per-case code location, rank of the first correct hypothesis, and event chain usefulness. Code location is the finest granularity at which any of the top three hypotheses matches the gold patch.}
\label{tab:per_case}
\footnotesize
\begin{tabular}{@{}l l l c c c c c c@{}}
\hline
 & \multicolumn{2}{c}{Code location} & \multicolumn{4}{c}{First correct hypothesis} & \multicolumn{2}{c@{}}{Usefulness} \\
Case & Workflow & Baseline & E1 & E2 & J1 & J2 & E1 & E2 \\
\hline
C01 & file     & -     & 2  & -- & -- & 2  & 2 & 2 \\
C02 & file     & -     & 1  & 3  & 3  & 3  & 3 & 3 \\
C03 & file     & file     & -- & -- & -- & -- & 3 & 3 \\
C04 & function & function & 1  & -- & -- & 1  & 2 & 2 \\
C05 & module   & file     & -- & -- & -- & -- & 3 & 3 \\
C06 & function & function & 1  & -- & 1  & 1  & 3 & 3 \\
C07 & function & function & 1  & -- & 1  & 1  & 5 & 4 \\
C08 & -     & module   & 1  & 1  & -- & 1  & 5 & 3 \\
C09 & function & file     & 2  & 2  & 2  & 2  & 5 & 4 \\
C10 & function & -     & 1  & -- & -- & 1  & 4 & 2 \\
C11 & function & function & 1  & 1  & 1  & 1  & 4 & 5 \\
C12 & file     & file     & -- & -- & -- & -- & 2 & 3 \\
\hline
\multicolumn{3}{@{}r}{\emph{solved of 12}} & 9 & 4 & 5 & 9 & & \\
\hline
\end{tabular}
\end{table}

\subsection{RQ2: Hypothesis Correctness and Event Chains}

\begin{table}[t]
\centering
\caption{Rater agreement and event chain ratings. $\alpha$ is ordinal Krippendorff's alpha over individual ratings; Solved counts the cases on which a pair agrees whether any top-3 hypothesis is correct. Event chain columns count ratings at level 3 or better out of 12 and at level 1.}
\label{tab:oracle_agreement}
\footnotesize
\begin{tabular}{@{}l r r r | r r@{}}
\hline
& \multicolumn{3}{c|}{Correctness} & Useful. & Clutter \\
Pair & $\alpha$ & Exact & Solved & $\alpha$ & $\alpha$ \\
\hline
E1--E2 & 0.33 & 0.31 & 7/12 & 0.58 & 0.44 \\
E1--J1 & 0.41 & 0.25 & 8/12 & 0.34 & -0.14 \\
E1--J2 & 0.63 & 0.28 & 12/12 & 0.35 & -0.51 \\
E2--J1 & 0.45 & 0.39 & 9/12 & 0.18 & 0.02 \\
E2--J2 & 0.33 & 0.31 & 7/12 & -0.07 & -0.43 \\
\hline
& \multicolumn{3}{c|}{Usefulness} & \multicolumn{2}{c}{Clutter} \\
Rater & Med. & $\geq3$ & $=1$ & Med. & $\geq3$ \\
\hline
E1 & 3.0 & 9 & 0 & 3.0 & 9 \\
E2 & 3.0 & 9 & 0 & 3.0 & 11 \\
J1 & 3.0 & 12 & 0 & 4.0 & 12 \\
J2 & 4.5 & 12 & 0 & 4.0 & 12 \\
\hline
\end{tabular}
\end{table}

We denote experts E1 and E2, and LLM judges GPT-5.6 Terra and Claude Sonnet 5 as J1 and J2. Table~\ref{tab:per_case} lists the rank of each rater's first hypothesis rated 4 or higher per case, with a dash if none reached 4. At least one expert accepted one of the three top-ranked hypotheses as naming the defect in 9 of 12 cases; both experts accepted in 4 of 12 cases. At level 3, which identifies the faulting subsystem but not the defect, both experts agreed in 8 cases.

The five disputed cases reflect a difference in rater strictness. E2's accepted set is a strict subset of E1's. E2 accepted nothing E1 rejected and rated the 36 hypotheses 0.47 points lower on average. Ordinal $\alpha$ on correctness is 0.33 (Table~\ref{tab:oracle_agreement}), which reflects that offset.

After rating, we compared the ranking with random order of each case's three hypotheses, which puts the first correct one at rank 1 with probability $k/3$ and at expected rank $4/(k+1)$ when $k$ of them are correct. E1's first correct hypothesis was at rank 1 in 7 of 9 solved cases, against 4.7 expected under random order, and at mean rank 1.22 against 1.67. E2's ranks matched random order (2 of 4 at rank 1, mean 1.75), though E2 accepted all three in C11, making order irrelevant. The workflow’s internal ranking scores are not comparable between cases. Over all 36 hypotheses, their Spearman correlation with mean expert correctness is 0.02.

No event chain was judged misleading, as none of 48 ratings chose usefulness level 1, ``\textit{following this would have sent me the wrong way}''. Each expert rated the chain as narrowing the search to the right area or better in 9 of 12 cases, both in 8, and the LLM judges were more generous throughout (Table~\ref{tab:oracle_agreement}).

The LLM judges rarely matched experts' hypothesis scores but often agreed on solved cases. J2 matched E1's score on 28\% of the hypotheses, yet marked the same 9 cases as solved. J1 agreed with the stricter expert E2 on 9 of 12 cases (Table~\ref{tab:oracle_agreement}).

Code-location accuracy and expert correctness agree in 8 of 12 cases (Table~\ref{tab:per_case}) and diverge in both directions. In C03, C05, and C12 the workflow named a file or module fixed in the gold patch, but no expert accepted any of its hypotheses, because their mechanisms were wrong. In C03 and C05 both experts still rated event chain usefulness 3, meaning the chain narrowed the search to the right area. In C08 no hypothesis matched the gold patch at any granularity, yet both experts rated the first-ranked hypothesis 4. The fix spans 17 files, which path matching captures poorly. The overall counts are similar, 10 of 12 at file level against 9 accepted, but cover different cases.

\begin{figure}[t]
    \centering
    \includegraphics[width=\linewidth]{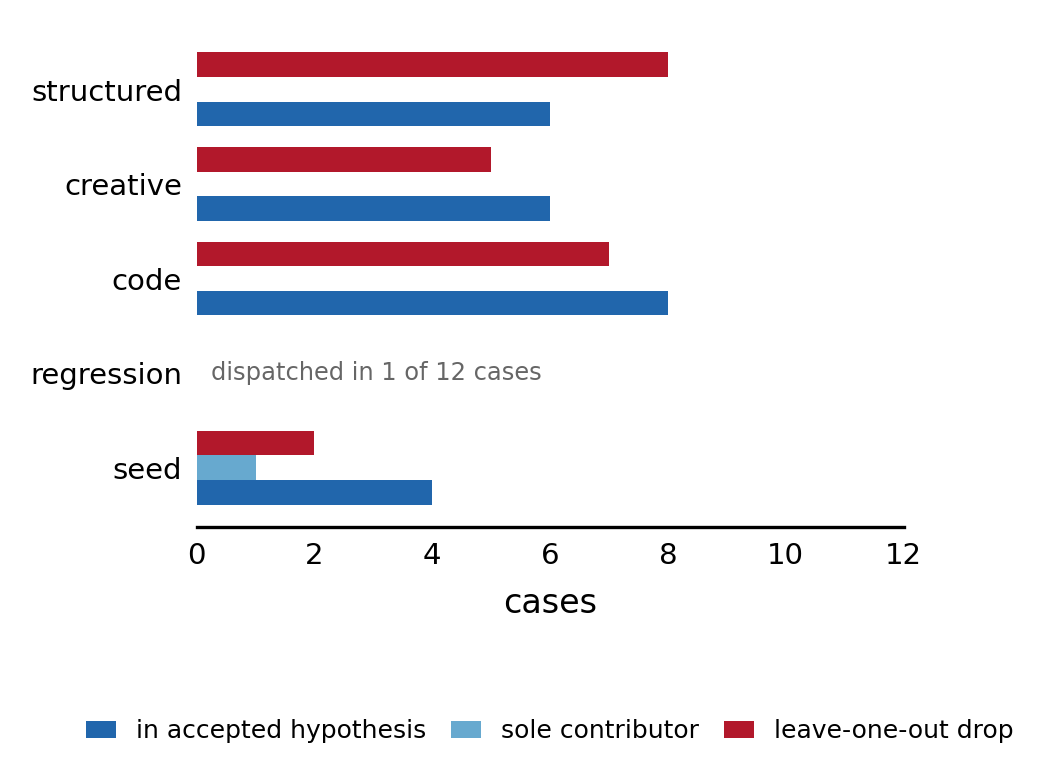}
    \caption{LLM analyst contribution to the nine accepted hypotheses, and the cases in which removing an analyst's hypothesis would degrade the best-of-3 code location.}
    \label{fig:rq3_contribution}
\end{figure}

\begin{figure}[t]
    \centering
    \includegraphics[width=\linewidth]{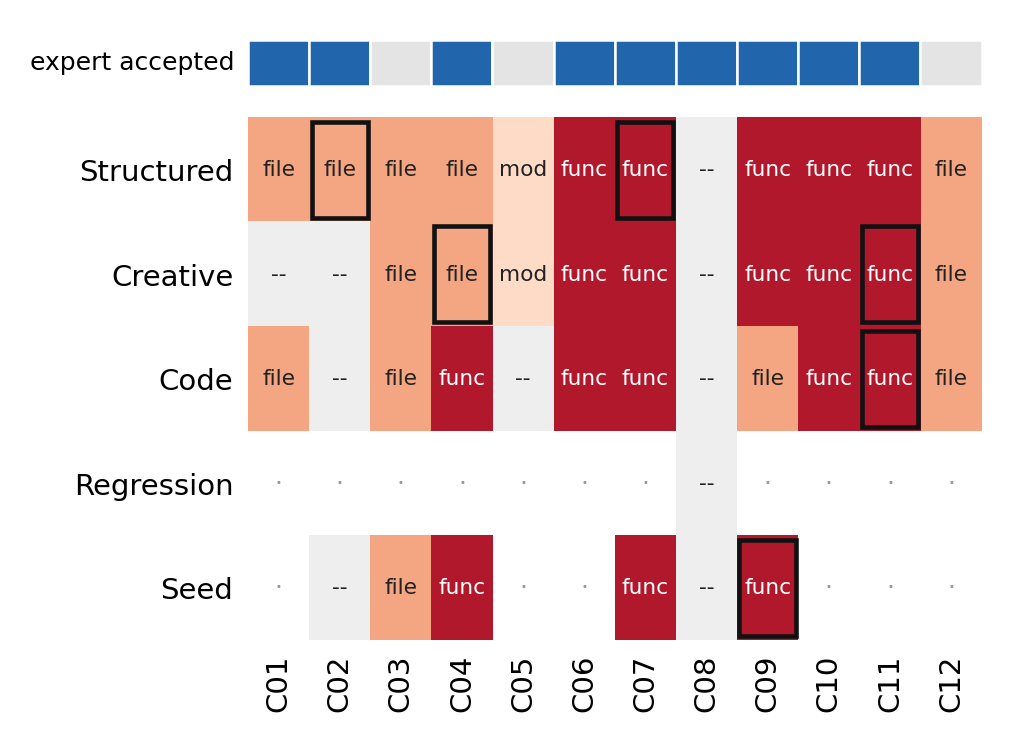}
    \caption{Deepest gold-patch granularity reached by any hypothesis each LLM analyst co-authored, per case. Boxed cells mark the analyst as the case's sole source of a matching location, dots mark cases where the analyst was not dispatched or abstained, and the top band marks the cases an expert accepted.}
    \label{fig:rq3_code_location}
\end{figure}

\subsection{RQ3: LLM analyst Contribution and Seed Handling}

Across the nine cases with an accepted hypothesis, the code analyst appears in the provenance of 8, the structured and creative analysts of 6 each, and the seed evaluator of 4 (Fig.~\ref{fig:rq3_contribution}). Only one accepted hypothesis had a single analyst, eight had two or more. Code location across all 12 cases ranks analysts differently (Fig.~\ref{fig:rq3_code_location}). There the structured analyst is strongest, and removing its hypotheses would lower the best code location in 8 of 12 cases, against 7 for the code analyst and 5 for the creative one. No analyst leads on both measures.

The regression analyst ran in all 12 cases but found regression evidence in only one and abstained in the other 11. Its single hypothesis was not accepted and reached no code location. The seed evaluator was dispatched in 6 of 12 cases, and its hypothesis always survived synthesis into the top three. In C09 it was the sole source of a hypothesis that both experts rated 5 (Fig.~\ref{fig:rq3_contribution}), the only sole-source hypothesis among the nine accepted hypotheses counted above. Isolating the reporter's theory therefore did not remove the theory from the final result.

\subsection{Seed Handling in Two Contrasting Cases}\label{sec:seedcases}
Two of six tickets with a seed show how the workflow handles a theory inconsistent with the fix.

In C02, the human reporter's theory named the right mechanism in the wrong component. The baseline followed this theory and missed the defect. The workflow found it but ranked the matching hypothesis third. In this case a truck under a reduced speed limit stopped at a ramp edge representing a forbidden drop-off, and in some repeated runs drove onto it. After seeing the speed briefly exceed the limit, the human reporter blamed missing hysteresis in the speed-limit control. The fix adds hysteresis, but elsewhere, to a gate that enables the search for non-drivable terrain only below a speed threshold. The brief overspeed switched this search off and hid the edge. All three baseline hypotheses followed the human reporter, treating the stop as a braking problem such as weak downhill cruise-control gains, and none matched the fix's code locations. In the workflow, the seed evaluator judged the reporter's claim weakened. The neutral description shared with the other LLM analysts kept the overspeed but dropped the claim that speed control caused it. Here the structured analyst, which never saw the seed, was the only one to trace the overspeed to the suppressed search. All four raters scored this hypothesis 4 or 5, and it held the case's only matching code locations, but it was ranked third. E1 also accepted the first-ranked hypothesis, which matched the fix nowhere, so the case is solved at rank 1 by the ratings and at rank 3 by code location.

In C07, the human reporter's theory pointed away from the defect. Both configurations reached the faulty function, but only the workflow's first-ranked hypothesis included the restart exposing the defect. A rollover-prevention test tilted the trailer on a ramp while the dump body was raised, until a roll-angle diagnostic stopped the mission. After a remote restart, the truck completed the dump with the tilt unchanged. The reporter suspected the test's static unload spot. In the neutral rewrite the restart was kept as a symptom but this theory was dropped. The seed evaluator, while falsifying the reporter's theory, found the real defect in tracing the restart: the restarted task creates a fresh roll-angle guard that reports a safe angle until its debounce timer expires, and the dump finishes within that window. Three LLM analysts without seed access reached the same mechanism, so the merged hypothesis ranked first and matched the fix at function level. E1 rated it 4 and E2 3, the LLM judges rated it 4 and 5. Unlike in C02, the baseline did not simply adopt the human reporter's theory this time. Its first hypothesis matched the same function but explained the dump through the roll-angle limits alone, without the restart, and only its third took up the reporter's theory.

\section{Discussion and Lessons Learned}\label{sec:discussion}
Our results from the twelve cases from one codebase support no general claims, but suggest lessons for teams building LLM based RCA from tickets and single recordings.

\subsection{Decomposition Improved Breadth and Ranking}
Within its top three hypotheses, the workflow matched two more cases than the baseline at every code granularity, and two to three more at rank 1. Both used the same model, tools, and repository snapshot, so they differ mainly in orchestration, but also in budget, prompting, and output schema. With 12 cases, a two-case gap may also reflect run-to-run variation.

Much of the difference can be seen in ranking. The baseline's second hypothesis matched the gold patch more often than its first, so it often failed to put its best explanation on top. The workflow ranks a hypothesis higher when several LLM analysts reach it independently, and eight of the nine accepted hypotheses had two or more contributors. In C07, three analysts converged on the defect and placed it first. The same rule demotes a correct hypothesis found by only one analyst, like in C02, where it ended up third. The workflow cost a median of 8.5 times as much as the baseline, though both costs are small compared with an engineer's time.

\subsection{Isolating the Human Reporter's Theory}
The baseline's handling of C02 shows the risk that motivated bias-isolated intake. All three baseline hypotheses adopted the human reporter's speed-control explanation, although the baseline could query the same telemetry. The workflow's structured analyst, however, ran in a separate agent session with this explanation isolated from its context, and could identify the correct cause. On the other hand, always discarding the reporter's theory would have lost the seed evaluator's hypothesis in C09, which both experts rated 5 and no other LLM analyst produced, so the isolation approach of our workflow was successful in both cases.

Two details of how we handle this isolation step proved important in practice. First, when rewriting the ticket for the main analysts, the intake step must keep factual observations (such as the brief overspeed in C02) while removing the reporter's guess about why they mattered. Without that overspeed clue, the structured analyst could not have traced the bug to the terrain check. Second, requiring the seed evaluator to actively look for counter-evidence helped it pivot from a wrong guess to the real bug during C07's mission restart. Although our 12 cases contain only one clear instance where the baseline was completely anchored by a bad hint (C02), isolating the hint costs only one additional agent run on tickets that include a guess and avoids throwing away useful operator insights (as in C09). Separating observed symptoms from suspected causes at intake is a simple pattern that can be reused in other bug-triage workflows.

\subsection{The Event Chain Holds Value on Its Own}
No rater judged an event chain misleading in any of the 48 ratings, and chains retained value when hypotheses failed. In C03 and C05 no expert accepted any hypothesis, yet both experts rated the chain as narrowing the search to the right area. Intermediate evidence should therefore be saved and presented as a result in its own right, not only as input to the analysts.

\subsection{Rating Causal Explanations Is Difficult}
Judging a free-text diagnosis requires understanding both the explanation and the fix, and each of our three measures falls short in this respect. Code-location accuracy is cheap and objective, yet it disagreed with the experts in 4 of 12 cases. The experts themselves, applying the same anchors to the same outputs, accepted a hypothesis in 9 and 4 cases, respectively. A study with a single expert would have reported either number with no sign of the other. The LLM judges, finally, matched the experts' case-level decisions but rarely their individual scores.

\section{Limitations}
\subsection{System Limits}
The workflow analyzes one recording per incident, which means unrecorded evidence, such as data from other site vehicles, reaches LLM analysts only through the ticket. Intake depends on an LLM to separate observation from theory, and a theory that the human reporter phrases as an observation could pass into the shared description unchanged.

The central extraction step determines what analysts see first. It flags deviations from expected behavior, so a cause that leaves no anomaly, such as a missing message or a value that stays within its usual range, can be absent from the event chain. The additions that LLM analysts propose to the event chain are tagged by origin but not assessed, so an analyst with weak hypotheses can still add clutter. Extraction rules and domain context files are also adapted to a specific setting and must be rewritten for other systems.

Evaluation showed that the workflow's internal ranking orders hypotheses within an incident but does not express correctness probability (Section~\ref{sec:results}). The corroboration signal used during ranking favors mechanisms reached by several LLM analysts and can work against a correct theory that was only found by one agent, which is why in C02 it placed the hypothesis matching the fix third.
\subsection{Threats to Validity}
\emph{Internal validity.} Raters coded with the gold patch visible, and as members of the development team they may already have worked on the defects. Familiarity with the defects and knowledge of the fix expose the event chain usefulness and clutter ratings to hindsight bias. The Claude judge uses the same model that generated the hypotheses it rates and is the more generous judge. While the GPT judge offers a cross-family comparison it cannot rule out self-preference.
\emph{External validity.} The cases come from one AHS codebase at one company, and all runs use one model. Other stacks, telemetry formats, or models may change absolute results and the workflow--baseline difference.

\emph{Construct validity.} Correctness uses one gold patch although other valid fixes may exist, and function-level matching only approximates the faulty region. The experts did not rate baseline outputs, so RQ1 rests only on code location. Credit ratios transfer to other providers only as far as their relative token prices agree.

\section{Conclusion}

We presented a multi-agent workflow that uses large language models to investigate software failures in autonomous heavy vehicles. Starting from an incident ticket and a vehicle recording, it builds an event chain describing what happened and produces ranked hypotheses with the code locations they point to. Several specialist agents examine the same evidence using different analysis methods, while a separate agent checks the human reporter's suspected cause. On 12 industrial incidents, the workflow identified the faulty file among its three highest-ranked hypotheses in 10 cases, compared with 8 for a single agent using the same model and tools, although the workflow cost more to run.

Our study suggests four lessons for building and evaluating such systems. First, using several LLM analysts equipped with different RCA techniques (algorithmic diversity) to examine an incident can help them find explanations that one agent misses. Agreement between LLM analysts can help rank these explanations, but it may also push a correct hypothesis down the list when only one analyst finds it. Second, a human reporter's suspected cause should be checked separately from the shared description of what happened. This lets the other LLM analysts explore alternatives while keeping the reporter's knowledge available to the investigation. Third, the event chain has value even when the proposed causes are wrong: it can still point engineers toward the right part of the system and help them understand how the failure developed. Finally, evaluation should distinguish finding the faulty code from explaining why the failure happened. In three cases, the workflow pointed to a file or module changed by the fix, but both experts rejected its explanations. Code-location accuracy is therefore useful for measuring how well the workflow narrows the search, but it does not establish that the proposed cause is correct. Expert judgments also differed substantially, so reporting both assessments gives a clearer picture of the workflow's strengths and remaining limitations.

Future work includes exploring automatic scenario generation to test the proposed hypotheses; one could use the event chain and each hypothesis to generate scenarios, run them in simulation, and examine whether the results support or contradict the proposed cause. Another direction is to investigate calibrating LLM analysts' confidence estimates against hypothesis correctness on separate validation incidents. One can test whether using these calibrated estimates to weight LLM analyst contributions reduces the influence of uncertain agreement and better preserves well-supported hypotheses found by a single analyst. A dedicated clean-up agent could also remove redundant or unsupported event chain entries after analysis. Finally, further evaluation on more incidents and in live investigations is needed to determine whether the workflow helps engineers find the cause faster.

\section*{Data Availability}
The incident tickets, vehicle recordings, source code, and expert ratings
used in this study are proprietary to Volvo Autonomous Solutions and
contain confidential information about a commercially deployed safety-critical
system. They therefore cannot be made publicly available. The workflow
design, analyst roles, hypothesis schema, and evaluation protocol are
described in Sections~IV--VI to support reimplementation in other settings.

\section*{The use of Generative AI}
The authors used Claude Opus~5.5 and Claude Sonnet~5.5 (Anthropic) to edit the manuscript's language. The authors also used GitHub Copilot with Claude Opus 5 and GPT-5 to help implement the telemetry extraction tools, artifact store, and evaluation
scripts, which the authors designed, reviewed, and tested. Apart from this and the LLM agents of the studied workflow, generative AI was not used to generate research content, results, or interpretations. The authors take full responsibility for the content.

\bibliographystyle{IEEEtran}
\bibliography{literature}
\end{document}